\documentclass{camera}
\usepackage{graphicx}

\def\kaos{{\sc Kaos}\@}
\graphicspath{{figures/}}

\begin{document}

\procinfo{Conference Proceedings Vol. 99}{``XLVII International Winter Meeting on Nuclear Physics''}{R.A. Ricci, W. K\"uhn and A. Tarantola (Eds.)\\SIF, Bologna, 2010}

\title{Strangeness physics with \kaos\ at MAMI}

\author{P. Achenbach {\normalfont on behalf of the} A1 Collaboration\thanks{
P.~Achenbach$^1$,
C.~Ayerbe Gayoso$^1$,
J.~C.~Bernauer$^1$,
R.~B\"ohm$^1$,
M.~B\"osz$^1$,
D.~Bosnar$^2$,
L.~Debenjak$^3$,
A.~Esser$^1$,
M.~O.~Distler$^1$,
M.~{G\'omez Rodr\'iguez}$^1$,
K.~Grie{\ss}inger$^1$,
P.~Jennewein$^1$,
M.~Makek$^2$,
H.~Merkel$^1$,
U.~M\"uller$^1$,
L.~Nungesser$^1$,
J.~Pochodzalla$^1$,
M.~Potokar$^3$,
S.~{S\'anchez Majos}$^1$,
B.~S.~Schlimme$^1$,
S.~\v Sirca$^{3,4}$,
Th.~Walcher$^1$,
M.~Wein\-riefer$^1$, and 
C. J.~Yoon$^1$
--$^1$Institut f\"ur Kernphysik, Johannes Gutenberg-Universit\"at
    Mainz, Germany;
--$^2$Department of Physics,
    University of Zagreb, Croatia;
--$^3$Jo\v zef
    Stefan Institute, Ljubljana, Slovenia;
--$^4$University of Ljubljana, Slovenia.}}

\organization{Institut f\"ur Kernphysik, J. J. Becherweg 45,\\ 
	Gutenberg-University, D-55099 Mainz, Germany\\
  E-mail: patrick@kph.uni-mainz.de\\[10mm] 
  }  

\maketitle

\begin{abstract}
  At the Institut f\"ur Kernphysik in Mainz, Germany, the microtron
  MAMI has been upgraded to 1.5\,GeV electron beam energy. 
  The magnetic spectrometer \kaos\ is now operated by the A1 
  collaboration to study strangeness electro-production. Its
  compact design and its capability to detect negative and positive
  charged particles simultaneously under forward scattering angles
  complements the existing spectrometers.
  In 2008 kaon production off a liquid hydrogen target was measured at 
  $\langle Q^2\rangle =$ 0.050\,(GeV$/c$)$^2$
  and 0.036\,(GeV$/c$)$^2$. Associated $\Lambda$ and $\Sigma^0$ hyperons were
  identified in the missing mass spectra. Major
  modifications to the beam-line are under construction and a new electron
  arm focal-surface detector system was built in order to use \kaos\ as a 
  double-arm spectrometer under zero degree scattering angle.
\end{abstract}
 
\pagestyle{plain}
\section{Introduction}
%
At the Institut f\"ur Kernphysik in Mainz, Germany, the microtron MAMI
has been upgraded to 1.5\,GeV electron beam energy~\cite{Kaiser2008}.  The microtron
delivers an electron beam with excellent
spatial and energy definition that can now be used
to study strange hadronic systems produced on solid-state or liquid cryogenic
targets. 

The elementary electro-production of kaons off the proton is used
mainly as a test of production mechanisms.  
Effective field theories expressed in terms of resonant baryon
formation and kaon exchange~\cite{Bennhold1999}
have been successfully applied to describe the data. However, in contrast to the well studied flavour $SU(2)$ sector, 
more precision data in the
threshold energy region are needed. Also for calculating hypernuclear
production cross-sections, precise data from very forward angle kaon production 
are urgently required.  In this kinematics the elementary amplitude serves as the basic
input, which determines the accuracy of predictions for
hypernuclei~\cite{Bydzovsky2006}.

For the strangeness programme
the \kaos\ spectrometer was recently dismantled at the SIS facility at
GSI and re-installed in the spectrometer hall at MAMI. \kaos\ is a very
compact magnetic spectrometer suitable especially for the detection of
kaons~\cite{Senger1993}. The first-order
focusing is achieved in the set-up at MAMI with a bending of the
central trajectories on both sides by $\sim$ 45$^\circ$ with a
momentum dispersion of 2.2\,cm$/$\% for the hadron arm and
4.0\,cm$/$\% for the electron arm. From the magnet optics, the beam diameter, and the spatial resolution in the focal-surface 
a first order momentum resolution of $\Delta p/p \sim$ 10$^{-3}$
is expected.

\section{Particle tracking and identification with Kaos}
%
The tracking of particles through \kaos\ is performed by means of two
large MWPC with a total of 2 $\times$ 310 analogue channels.  Five
cathode wires are connected together and are brought to one charge
sensitive pre-amplifier followed by an ADC card.  The transputer-based
read-out system is connected to a multi-link card of a front-end computer.
To determine the particle track the measured charge
distributions are analysed by the centre-of-gravity method.

Particle identification in the \kaos\ spectrometer is based on the
particle's time-of-flight and its specific energy loss. A segmented
scintillator wall with 30 paddles read out at both ends by
fast photomultipliers is located near the focal-surface and measures the
arrival time. A second wall with 30 paddles is used to
discriminate valid tracks against background events. A top--bottom
mean timing for deriving the trigger is performed by summing the
analogue signals. The signal amplitudes were corrected for the particle's path-length through the scintillator bulk material and the light absorption inside the paddle.  

\begin{figure}
  \centering
  \includegraphics[width=0.7\textwidth]{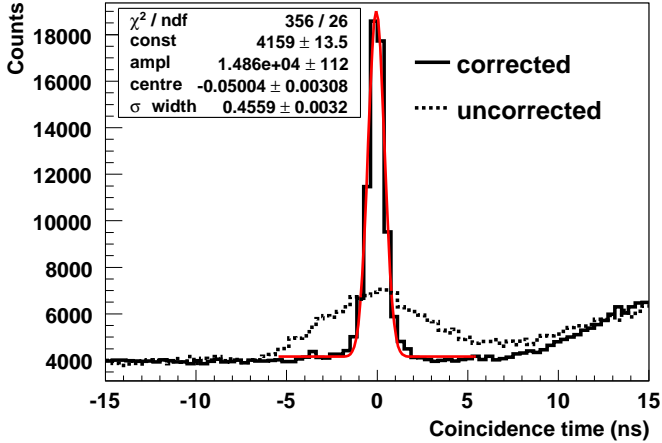}
  \caption{Coincidence time spectrum for the p$(e,e'\pi)$ reaction
    before and after corrections for the flight-time and trigger-time
    jitter. A Gaussian distribution on top of a constant background
    was fitted to the corrected spectrum. The width of the $(e',\pi)$
    peak is $\Delta t_{\it FWHM} =$ 1.07\,ns, which is a typical
    inter-spectrometer time resolution.}
  \label{fig:TOF}
\end{figure}

The time spectrum is systematically broadened by the propagation time
dispersion inside the scintillator, the time differences between
different scintillator paddles and their associated electronic
channels, and by the variation of the time-of-flight, being
proportional to the path-length through the spectrometer.  The
coincidence time spectrum for the p$(e,e'\pi)$ reaction before and
after corrections is shown in fig.~\ref{fig:TOF}, where the
flight time, $t$, was corrected by using the reconstructed momentum,
$p$, and path-length, $L$, under the assumption
that a pion was detected. The Gaussian width of the
$(e',\pi)$ peak is $\Delta t_{\it FWHM} =$ 1.07\,ns, which is a
typical inter-spectrometer time resolution.

\begin{figure}
  \centering
  \includegraphics[width=0.7\textwidth]{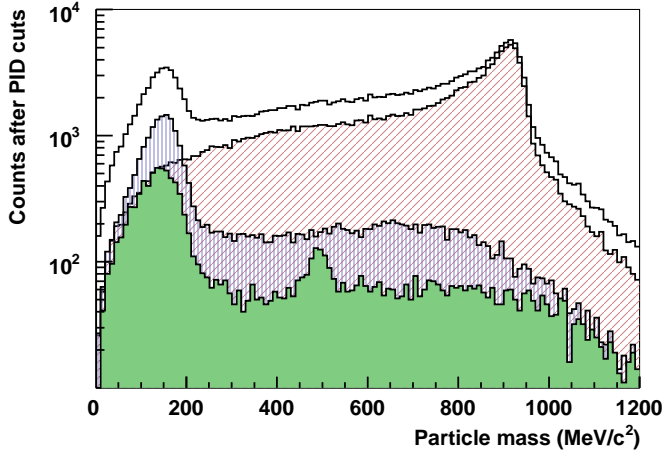}
  \caption{Mass distribution of the detected particles based on
    time-of-flight measurement and the reconstructed momentum. The
    outer contour includes all events with valid tracks in the MWPC,
    the underlying curves show event samples that were cut on the
    specific energy-loss expectation for protons (red, leaning hatch),
    pions (blue, vertical hatch), and kaons (green, solid fill).}
  \label{fig:mass}
\end{figure}

The necessity of the pion and proton
suppression is seen in fig.~\ref{fig:mass} where the mass distribution
of the detected particles, $ M = p/(c\cdot\beta\gamma) = p/(L t) \sqrt{1-(L t/c)^2}$, is shown under the effect of specific 
energy-loss cuts. Kaons were identified subsequently by a
cut on the coincidence time.

\section{Pilot kaon electro-production measurements}
%
A pilot experiment on the electro-production of kaons off a liquid hydrogen target was performed with an electron beam of 1.508\,GeV energy in 2008. The
reaction can lead to two possible final states with either a $\Lambda$ or
$\Sigma^0$ hyperon, which are easily separable by a missing mass
analysis. The data was taken at two different kinematic settings in
$(e,e'K)$ reactions with kaons in the momentum range of
400--700\,MeV$/c$ and angular range of 21--43$^\circ$.  Positive kaons
were detected in \kaos\ in coincidence with the scattered electron
into spectrometer~B.  The electron was identified by its minimum
ionisation in the scintillators of spectrometer~B and a signal in the
gas \v{C}erenkov detector.  The momentum transfer squared
was $\langle Q^2\rangle =$ 0.050\,(GeV$/c$)$^2$,
resp.\ 0.036\,(GeV$/c$)$^2$, and the total
energy in the virtual-photon-nucleon cm system
was $\langle W\rangle =$
1.670\,GeV, resp.\ 1.750\,GeV. The kinematic conditions for two
beam-times with a total integrated luminosity of 284\,fbarn$^{-1}$ taken on a 48\,mm $\ell$H$_2$ target with 1--4\,$\mu$A beam current are
summarised in table~\ref{tab:kinematics}.

\begin{table}
  \caption{Experimental settings during
    the kaon electro-production beam-times of 2008. The second setting
    was selected to acquire data from $\Lambda$ and $\Sigma^0$ production
    channels, with the two different associated kaon momenta being 
    simultaneously within the large momentum acceptance of \kaos.}
    {\begin{tabular}{@{}cccccccc@{}} 
    \hline
      \multicolumn{4}{c}{virt.\ photon} 
      & \multicolumn{2}{c}{electron arm}
      & \multicolumn{2}{c}{kaon arm} \\
    \hline\\[-4mm]
    $\langle Q^2 \rangle$ & $\langle W \rangle$ & $\langle \epsilon \rangle$ 
    & $\langle \omega \rangle$
    & $\langle q^{lab}_{e'} \rangle$ & $\langle \theta^{lab}_{e'} \rangle$
    & $\langle p^{lab}_K \rangle$
    & $\langle \theta^{lab}_K \rangle$ \\
    (GeV$/c$)$^2$ & GeV & (trans.) & GeV
    & GeV$/c$ & deg & GeV$/c$ & deg \\
    \hline
    0.050 & 1.670 & 0.540 & 1.044 & 0.455 & 15.8 
    & $\Lambda:$ 0.466 & -31.5 \\
    0.036 & 1.750 & 0.395 & 1.182 & 0.318 & 15.5 
    & $\Lambda:$ 0.642 & -31.5 \\
    &       &       &       &       &       &      
    $\Sigma^0:$  0.466 & \\
    \hline
  \end{tabular}
  \label{tab:kinematics}}
\end{table}

After electron and kaon identification, the measured momenta allow for
a full reconstruction of the missing energy and missing momentum of
the recoiling system.  The missing mass $M_X$ is calculated from the
four-momenta $q^\mu$ of the virtual photon and the four-momentum
$p_K^\mu$ of the detected kaon according to $ M^2_X = (q^\mu +
P^\mu_{targ} - p^\mu_K)^2$, where $P^\mu_{targ} = (M_{targ}, \vec{0})$
is the target four-momentum.

\begin{figure}
    \includegraphics[width=0.49\textwidth]{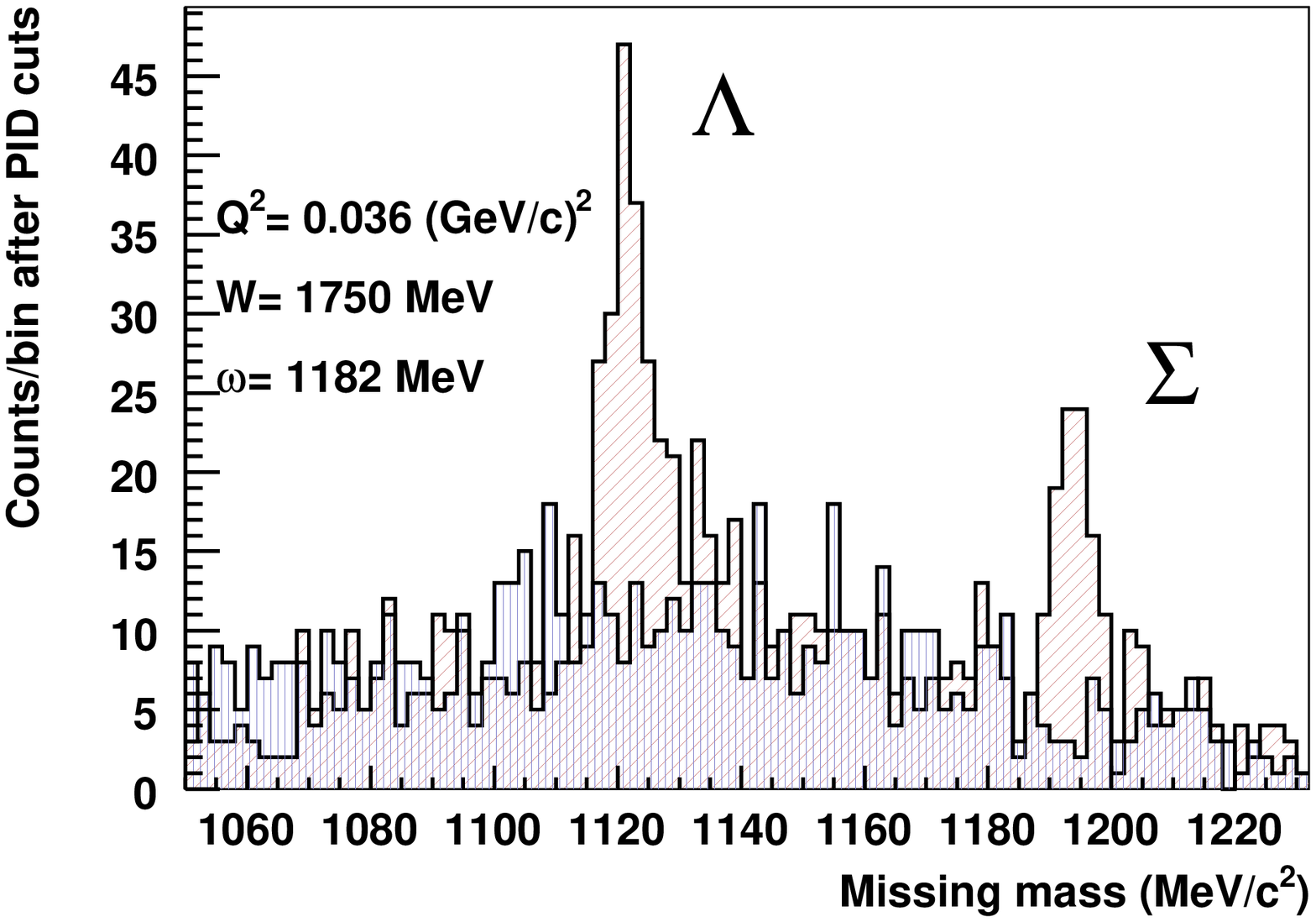}
    \includegraphics[width=0.49\textwidth]{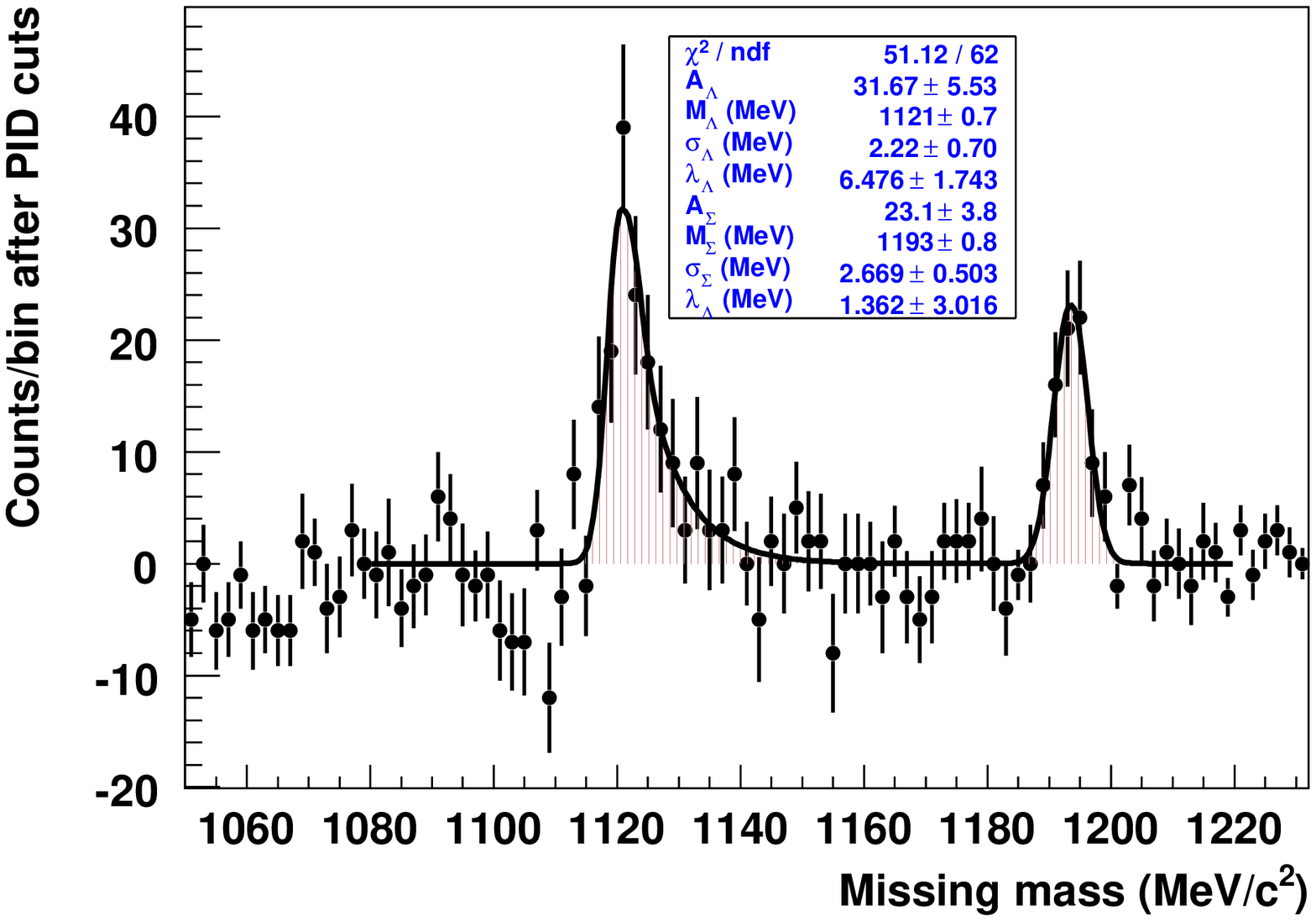}
  \caption{Preliminary missing mass spectra in the p$(e,e'K^+)$Y
    reaction at the $Q^2 =$ 0.036\,(GeV$/c$)$^2$ kinematical point as defined in the
    text. The random background distribution is shown as a blue histogram in the left figure. The right figure shows the background subtracted spectrum with a fit to the two peaks. The shape of each peak was assumed to include a Gaussian and an exponential tail.}
  \label{fig:mm}
\end{figure}

An example of the preliminary missing mass spectra is shown in
fig.~\ref{fig:mm} which demonstrate the power of the spectrometer
facility to detect open strangeness channels. The mass
resolution is limited by errors in the estimated transfer matrix
that was not yet corrected. 
The overlaid blue histogram shows the random background distribution
in two averaged $(e',K)$ coincidence time side-bands with the
appropriate weights.

\section{Extension to hypernuclei electro-production}
The electro-production of hypernuclei offers the unique possibility to
vary the energy and momentum transfer independently and to gain
information on hypernuclear wave-functions. Such experiments are
planned at MAMI for single $\Lambda$-hypernuclei in light
targets~\cite{Pochodzalla2005}.  The special kinematics for
electro-production of hypernuclei requires the detection of both, the
associated kaon and the scattered electron, at very forward laboratory
angles which will be achieved by instrumenting the \kaos\
spectrometer in two arms, to either side of
the main dipole. This implies that the electron beam must be
steered through the spectrometer and
a magnetic chicane comprising two compensating sector magnets is under
development. A new vacuum chamber has been installed and the spectrometer platform has been mechanically adapted to the two-arm operation.
A new coordinate detector has been developed for the
electron arm~\cite{Achenbach-HYP06}.  It consists of two vertical
planes of 18,432 fibres that will be supplemented by one or two
horizontal planes. Detectors and electronics for the 4,608 read-out and level-1 trigger channels are now being installed.

\section*{Acknowledgements}
We gratefully acknowledge the essential support from the technical
staff at Mainz and from the former KaoS
collaboration at GSI.

Work supported in part by Bundesministerium f{\"u}r Bildung und
For\-schung (bmb+f) under contract no.\ 06MZ176 and in part by the
Federal State of Rhineland-Palatinate and by the Deutsche
Forschungsgemeinschaft with the Collaborative Research Center 443.

%
\end{document}